\documentclass
[superscriptaddress,secnumarabic,amssymb,amsmath,nobibnotes,aps,prd,showkeys,nofootinbib,nopacsnumber,onecolumn,10pt]{revtex4}%
\usepackage{graphicx}
\usepackage{bm}
\usepackage{amsmath}
\usepackage{amsfonts}
\usepackage{amssymb}%
\providecommand{\U}[1]{\protect\rule{.1in}{.1in}}

\newcommand{\be}{\begin{equation}}
\newcommand{\ee}{\end{equation}}

\newcommand{\mincir}{\raise
-3.truept\hbox{\rlap{\hbox{$\sim$}}\raise4.truept\hbox{$<$}\ }}
\newcommand{\magcir}{\raise
-3.truept\hbox{\rlap{\hbox{$\sim$}}\raise4.truept\hbox{$>$}\ }}

\begin{document}
\title{Observational Constraints on Scalar Field--Matter Interaction in Weyl
Integrable Spacetime}
\author{Andronikos Paliathanasis}
\email{anpaliat@phys.uoa.gr}
\affiliation{Department of Mathematics, Faculty of Applied Sciences \& Institute of Systems
Science, Durban University of Technology, Durban 4000, South Africa}
\affiliation{Centre for Space Research, North-West University, Potchefstroom 2520, South Africa}
\affiliation{Departamento de Matem\`{a}ticas, Universidad Cat\`{o}lica del Norte, Avda.
Angamos 0610, Casilla 1280 Antofagasta, Chile}
\affiliation{National Institute for Theoretical and Computational Sciences (NITheCS), South Africa.}

\begin{abstract}
We test an analytic cosmological solution within the framework of Weyl
Integrable Spacetime using current observational data. In this model, dark
energy is described by a pressureless fluid, while a scalar field arises
naturally through the definition of the connection. This gravitational theory
reveals a Chameleon Mechanism leading to a nonzero interaction between the
scalar field and the matter sector. This model extends the standard $\Lambda
$CDM cosmology by introducing one additional degree of freedom, allowing for
deviations from $\Lambda$CDM dynamics. For the observational constraints we
consider the Supernova data of Pantheon+ collaboration, the Cosmic
Chronometers, the Baryonic Acoustic Oscillators of DESI DR2 collaboration and
the gamma-ray bursts. We find that Weyl Integrable Spacetime fits the data in
a better way than the $\Lambda$CDM. When all datasets are considered, the
statistical comparison indicates a weak to moderate preference for Weyl
Integrable Spacetime according to the Akaike Information Criterion and
Jeffrey's scale for the Bayesian evidence.

\end{abstract}
\date{\today}
\maketitle


\section{Introduction}

Weyl Integrable Spacetime (WIS) is a gravitational theory with nonzero
nonmetricity \cite{salim96,ww1,ww2,ww3,ww4,ww5}, which extends the concept of
General Relativity. The connection that defines the covariant derivative is
that of a conformally related metric, where the conformal factor contributes
dynamically to the gravitational theory \cite{salim96}. In WIS, the scalar
field is introduced from the connection without attributing higher-order
derivatives, as is the case in $f\left(  R\right)  $ gravity \cite{tomas}.
Gravitational theories with nonmetricity, where the degrees of freedom of the
connection lead to the introduction of a scalar field, include the
nonmetricity scalar-tensor theory \cite{nm1} and $f\left(  Q\right)  $ gravity
\cite{de05,y1,y2,y3}. A more general framework of nonmetricity-based
gravitational theories was recently introduced in \cite{lehel}; see also the
discussion in \cite{lehel1}.

In the case of vacuum, WIS is equivalent to General Relativity with a scalar
field, either quintessence
\cite{de04,qq1,qq2,qq3,qq4,qq5,qq6,qq7,qq8,qq9,q10,q11} or phantom fields
\cite{q14,q15,q16,q17,q18}. However, in the presence of a matter source,
interaction terms naturally emerge as a consequence of the geometric structure
of the theory. Moreover, when the matter source is an additional scalar field,
the Chiral-Quintom theory is recovered \cite{cq1,cq2}.

Gravitational models with dark energy-dark matter interaction have been widely
studied in the literature in order to address the coincidence problem or
cosmological tensions \cite{Amendola:1999er,con1,con2,ht3,ht4,ht5,ht6}. In
these models, there exists an energy transfer between dark energy and dark
matter. The mass of the scalar field depends on the energy density of the
matter source, allowing the scalar field to mimic the energy density of the
matter's behavior described as the Chameleon mechanism \cite{ch1,ch2}. The
majority of the interacting models in the literature are phenomenological
\cite{ss4,ss5,ss11,ss12,ss18}. Nevertheless, the geometric framework of WIS
provides a theoretical justification for the interaction term. An alternative
approach to theoretically construct scalar field-matter interaction is through
the conformal transformation that relates the Jordan and Einstein frames in
scalar-tensor theories \cite{sf3}. The dynamics of scalar field-matter
interacting models have been widely studied in the literature
\cite{dn1,dn2,dn3,dn4,dn5,dn6,dn7,dn8,dn9,dn10}. It was found that, in the
case of a phantom field, the presence of interaction is important in order for
the trajectories of the cosmological field equations to avoid solutions that
describe Big Rip singularities \cite{cop2}. Moreover, some exact and analytic
solutions in scalar field-matter interacting models were previously determined
in \cite{weylnoether,sh0}.

In this study, we focus on the analytic solution in WIS presented in
\cite{weylnoether}, which corresponds to an exponential potential, with the
matter source described by an ideal gas with a constant equation of state
parameter. We consider the case of dark matter and study the effects of the
free parameters on the behavior of the solution. We show that the essential
free parameters of this model are three. For the observational data analysis,
we make use of the recent release of Baryon Acoustic Oscillation (BAO) data by
the DESI DR2 collaboration, as well as gamma-ray bursts (GRBs). Both datasets
have been widely examined in the literature for the study of cosmological
models
\cite{d1,d2,d3,d4,d5,d6,d7,d8,d9,d10,d11,d12,d13,d14,d15,d16,d17,d18,d19,d20,d21,d22,d23,d24,d25,d26,d27,d28,d29,d30,d31,d32}%
.

The structure of the paper is as follows.

In Section \ref{sec2} we introduce the basic properties and definitions of the
WIS. In this geometric framework, the scalar field is introduced naturally in
the gravitational field equations from the definition of the connection of the
conformally related geometry. Furthermore, there appears a nonzero interaction
term of the scalar field with the matter source related to the conformally
related connection. This reveals the Chameleon mechanism. The field equations
for this gravitational model are of second order, and in Section \ref{sec3} we
review a recent analytic solution determined in the literature. We discuss in
detail the nature of the integration constants in the analytic solutions
provided by the integration. We end with an analytic expression for the Hubble
function with three free parameters, the Hubble constant, the energy density
of the matter, and the coupling parameter which defines the interaction
between the scalar field and the matter. We demonstrate that when the latter
parameter reaches a limit, the $\Lambda$CDM universe is recovered.

Section \ref{sec4} presents the main results of this work, where we employ
cosmological observations in order to constrain the analytic cosmological
model of the WIS. Specifically, we consider the Supernova datasets of the
Pantheon+ collaboration, the observational Hubble parameter data as they are
measured by the Cosmic Chronometers, and the BAO data from the DESI DR2
collaboration. These datasets provide us with events for redshifts $z<2.4$.
Hence, we introduce the GRBs, which provide us with events for large values of
the redshifts $z<8.1$. We use the $\Lambda$CDM model as the baseline for
comparing the statistical parameters provided by the WIS datasets. Finally, in
Section \ref{sec5} we draw our conclusions.

\section{Weyl Integrable Spacetime}

\label{sec2}

In WIS, the fundamental geometric objects are the metric tensor $g_{\mu\nu}$
and the covariant derivative $\tilde{\nabla}{\mu}$, defined by the symmetric
connection $\tilde{\Gamma}{\mu\nu}^{\kappa}$, which differs from the
Levi-Civita connection, such that \cite{salim96}%
\begin{equation}
\tilde{\nabla}_{\kappa}g_{\mu\nu}=\phi_{,\kappa}g_{\mu\nu}. \label{c.01}%
\end{equation}
Consequently, $\tilde{\Gamma}_{\mu\nu}^{\kappa}$ is defined as
\begin{equation}
\tilde{\Gamma}_{\mu\nu}^{\kappa}=\Gamma_{\mu\nu}^{\kappa}+Q_{\mu\nu}^{\kappa},
\label{c.02}%
\end{equation}
where the connection $\Gamma_{\mu\nu}^{\kappa}$ defines the covariant
derivative $\nabla_{\kappa}$ and it coincides with the Levi-Civita connection
for metric $g_{\mu\nu}$, that is, $\nabla_{\kappa}g_{\mu\nu}=0$. The
nonmetricity component $Q_{\mu\nu}^{\kappa}$ reads
\begin{equation}
Q_{\mu\nu}^{\kappa}=-\phi_{,(\mu}\delta_{\nu)}^{\kappa}+\frac{1}{2}%
\phi^{,\kappa}g_{\mu\nu}. \label{c.03}%
\end{equation}
Hence, if $\tilde{g}_{\mu\nu}=\phi g_{\mu\nu}$ be a conformally related metric
to $g_{\mu\nu}$, with conformal factor the scalar field $\phi$, then by the
definition, (\ref{c.01}), (\ref{c.02}) and (\ref{c.03}) we conclude that
$\tilde{\Gamma}_{\mu\nu}^{\kappa}$ is equivalent to the Levi-Civita connection
for the conformal metric $\tilde{g}_{\mu\nu}\,$.

Within the WIS the Einstein-Hilbert Action is modified as \cite{salim96}
\begin{equation}
S_{W}=\int dx^{4}\sqrt{-g}\left(  \widetilde{R}+\xi\left(  \widetilde{\nabla
}_{\nu}\left(  \widetilde{\nabla}_{\mu}\phi\right)  \right)  g^{\mu\nu
}-V\left(  \phi\right)  +\mathcal{L}_{m}\right)  , \label{c.04}%
\end{equation}
where $\widetilde{R}$ is the Ricci scalar defined by the connection
$\tilde{\Gamma}_{\mu\nu}^{\kappa}$, that is$\frac{{}}{{}}$%
\begin{equation}
\tilde{\nabla}_{\nu}\left(  \tilde{\nabla}_{\mu}u_{\kappa}\right)
-\tilde{\nabla}_{\mu}\left(  \tilde{\nabla}_{\nu}u_{\kappa}\right)  =\tilde
{R}_{\kappa\lambda\mu\nu}u^{\lambda}. \label{c.05}%
\end{equation}
with Ricci tensor
\begin{equation}
\tilde{R}_{\mu\nu}=R_{\mu\nu}-\tilde{\nabla}_{\nu}\left(  \tilde{\nabla}_{\mu
}\phi\right)  -\frac{1}{2}\left(  \tilde{\nabla}_{\mu}\phi\right)  \left(
\tilde{\nabla}_{\nu}\phi\right)  -\frac{1}{2}g_{\mu\nu}\left(  \frac{1}%
{\sqrt{-g}}\tilde{\nabla}_{\nu}\tilde{\nabla}_{\mu}\left(  g^{\mu\nu}\sqrt
{-g}\phi\right)  -g^{\mu\nu}\left(  \tilde{\nabla}_{\mu}\phi\right)  \left(
\tilde{\nabla}_{\nu}\phi\right)  \right)  , \label{c.06}%
\end{equation}
and Ricci scalar%
\begin{equation}
\tilde{R}=R-\frac{3}{\sqrt{-g}}\tilde{\nabla}_{\nu}\tilde{\nabla}_{\mu}\left(
g^{\mu\nu}\sqrt{-g}\phi\right)  +\frac{3}{2}\left(  \tilde{\nabla}_{\mu}%
\phi\right)  \left(  \tilde{\nabla}_{\nu}\phi\right)  , \label{c.07}%
\end{equation}
Here, $R_{\mu\nu}$ and $R$ are the Ricci tensor and scalar, respectively,
defined for the original metric tensor $g_{\mu\nu}$. The parameter $\xi$ is an
arbitrary coupling constant introduced to ensure a nonzero contribution of the
scalar field's kinetic term in the field equations. The function $V\left(
\phi\right)  $ describes the mass of the scalar field, and $\mathcal{L}_{m}$
is the Lagrangian function for the matter source.

Where $\mathcal{L}_{m}$ describes a perfect fluid with energy density $\rho$
and pressure component $p$, then the gravitational field equations are
\begin{equation}
\widetilde{G}_{\mu\nu}=\tilde{T}_{\mu\nu}^{\phi}+\tilde{T}_{\mu\nu}^{m}
\label{c.08}%
\end{equation}
in which~$\widetilde{G}_{\mu\nu}$ is the Einstein tensor with respect the
metric $\widetilde{g}_{\mu\nu}$, $\tilde{T}_{\mu\nu}^{\phi}$ is the energy
momentum tensor which attributes the scalar field contribution \cite{salim96}%
\begin{equation}
T_{\mu\nu}^{\phi}=-\widetilde{\nabla}_{\nu}\left(  \widetilde{\nabla}_{\mu
}\phi\right)  +\left(  2\xi-1\right)  \left(  \widetilde{\nabla}_{\mu}%
\phi\right)  \left(  \widetilde{\nabla}_{\nu}\phi\right)  -\xi g_{\mu\nu
}g^{\kappa\lambda}\left(  \widetilde{\nabla}_{\kappa}\phi\right)  \left(
\widetilde{\nabla}_{\lambda}\phi\right)  -V\left(  \phi\right)  g_{\mu\nu}
\label{c.09}%
\end{equation}
and $\tilde{T}_{\mu\nu}^{m}$ is the energy-momentum tensor for the perfect
fluid, where in the $1+3$ decomposition with respect to the comoving observer
$u^{\mu}$ reads \cite{salim96}%
\begin{equation}
\tilde{T}_{\mu\nu}^{m}=\left(  \widetilde{\rho}_{m}+\widetilde{p}_{m}\right)
u_{\mu}u_{\nu}+\widetilde{p}g_{\mu\nu}. \label{c.10}%
\end{equation}
The new parameters $\widetilde{\rho}_{m},~\widetilde{p}_{m}\,\ $are the energy
density and pressure~components for the matter source multiplied by the factor
$e^{-\frac{\phi}{2}}$, that means $(\widetilde{\rho},~\widetilde
{p})=(e^{-\frac{\phi}{2}}\rho_{m},e^{-\frac{\phi}{2}}p_{m}).$

In terms of the Einstein tensor $G_{\mu\nu}$ for the metric tensor $g_{\mu\nu
}$, the field equations (\ref{c.08}) read%
\begin{equation}
G_{\mu\nu}=T_{\mu\nu}^{\phi}+\tilde{T}_{\mu\nu}^{m} \label{c.11}%
\end{equation}
where now
\begin{equation}
T_{\mu\nu}^{\phi}=\lambda\left(  \phi_{,\mu}\phi_{,\nu}-\frac{1}{2}g_{\mu\nu
}\phi^{,\kappa}\phi_{,\kappa}\right)  -V\left(  \phi\right)  g_{\mu\nu}
\label{c.12}%
\end{equation}
and the new parameter $\lambda$ is defined as $\lambda=2\xi-\frac{3}{2}$.

In the case of the vacuum, the Bianchi identity $\nabla_{\nu}G_{~}^{\mu\nu}=0$
leads to the conservation law $\nabla_{\nu}T_{~~~~~~~~}^{\phi~\mu\nu}=0$, that
is, \cite{salim96}
\begin{equation}
u^{\nu}\nabla_{\nu}\phi\left(  -\lambda g^{\mu\nu}\nabla_{\mu}\nabla_{\nu}%
\phi+V_{,\phi}\right)  =0, \label{c.14}%
\end{equation}
and the theory is equivalent with that of General Relativity with a
quintessence $\left(  \lambda>0\right)  $ scalar field or phantom field
$\left(  \lambda<0\right)  $.

However, in the presence of the matter source, the Bianchi identity gives%
\begin{equation}
\nabla_{\nu}\left(  T^{\phi~~\mu\nu}+\tilde{T}^{m~~\mu\nu}\right)  =0,
\label{c.15}%
\end{equation}
or equivalently,%
\begin{equation}
\nabla_{\nu}T^{\phi~~\mu\nu}=Q~,~\nabla_{\nu}\tilde{T}^{m~~\mu\nu}=-Q
\label{c.16}%
\end{equation}
where interaction function $Q$ is given by
\begin{equation}
Q=\frac{1}{2}\widetilde{\rho}\;u^{\nu}\nabla_{\nu}\phi. \label{c.17}%
\end{equation}

Indeed, the interaction between the scalar field and the matter source follows
as a geometric consequence of WIS. At this point, it is important to mention
that the interaction function $Q$ is proportional to the term $\phi_{,\nu}
u^{\nu}$. This is essential in order to avoid limitations when the scalar
field is constant.

Last but not least, expressions (\ref{c.16}) are written in the equivalent
form%
\begin{equation}
u^{\nu}\nabla_{\nu}\phi\left(  -\lambda g^{\mu\nu}\nabla_{\nu}\nabla_{\mu}%
\phi+V\left(  \phi\right)  +\frac{1}{2}e^{-\frac{\phi}{2}}\rho_{m}\right)  =0
\label{c.18}%
\end{equation}%
\begin{equation}
\left(  \nabla_{\mu}e^{-\phi}\rho_{m}\right)  u^{\mu}+e^{-\phi}\nabla_{\mu
}u^{\mu}\left(  \rho_{m}+p_{m}\right)  =0. \label{c.19}%
\end{equation}

\subsection{Chameleon Cosmology}

For a perfect fluid, the Lagrangian function $\mathcal{L}{m}$ is proportional
to the energy density, that is, $\mathcal{L}_{m}\sim\widetilde{\rho}_{m}$, or
equivalently, $\mathcal{L}_{m}\sim e^{-\frac{\phi}{2}}\rho_{m}$. Thus, the
gravitational field equations (\ref{c.11}) are derived from the variation of
the following equivalent Action Integral
\begin{equation}
S=\int\sqrt{-g}d^{4}x\left(  R-\frac{1}{2}g^{\mu\kappa}\nabla_{\mu}\phi
\nabla_{\kappa}\phi-V\left(  \phi\right)  -e^{-\frac{\phi}{2}}\rho_{m}\right)
.
\end{equation}
This gravitational Action Integral is known as Chameleon gravity
\cite{ch1,ch2}. Due to the nonzero interaction term between the scalar field
and matter, there exists an energy transfer between the two fluids,
proportional to the energy density of matter $\rho_{m}$. This interaction
allows the scalar field to acquire a large mass in dense environments, and a
small mass in regions of low energy density, such as on cosmic scales
\cite{ch1,ch2}.

Within the spatially flat FLRW universe with metric tensor
\begin{equation}
ds^{2}=-dt^{2}+a^{2}\left(  t\right)  \left(  dx^{2}+dy^{2}+dz^{2}\right)  ,
\end{equation}
and comoving observer $u^{\mu}=\delta_{t}^{\mu}$, $u^{\mu}u_{\mu}=-1$, the
cosmological field equations are%
\begin{equation}
3H^{2}-\frac{\lambda}{2N^{2}}\dot{\phi}^{2}-V\left(  \phi\right)
-e^{-\frac{\phi}{2}}\rho_{m}=0, \label{ww.17}%
\end{equation}%
\begin{equation}
\dot{H}+H^{2}+\frac{1}{6}e^{-\frac{\phi}{2}}\left(  \rho_{m}+3p_{m}\right)
+\frac{1}{3}\left(  \lambda\dot{\phi}^{2}-V\left(  \phi\right)  \right)  =0,
\label{ww.18}%
\end{equation}%
\begin{equation}
\ddot{\phi}+3H\dot{\phi}+V\left(  \phi\right)  +\frac{1}{2\lambda}%
e^{-\frac{\phi}{2}}\rho_{m}=0, \label{ww.19a}%
\end{equation}%
\begin{equation}
\dot{\rho}_{m}+3H\left(  \rho_{m}+p_{m}\right)  -\rho_{m}\dot{\phi}=0.
\label{ww.20a}%
\end{equation}
where, $H = \frac{\dot{a}}{a}$ is the Hubble function, which defines the
expansion rate $\nabla_{\mu} u^{\mu} = 3H$, where a dot denotes the total
derivative with respect to time, i.e., $\dot{a} = \frac{da}{dt}$. Moreover, we
have assumed that the scalar field and the matter source inherit the
symmetries of the FLRW metric; that is, they are homogeneous: $\phi= \phi(t)$
and $\rho_{m} = \rho_{m}(t)$.

Consider now the perfect fluid to be an ideal gas with constant equation of
state parameter $w_{m}=\frac{p_{m}}{\rho_{m}}$, then from (\ref{ww.20a}) it
follows~$\rho_{m}=\tilde{\rho}_{m0}a^{-3\left(  w_{m}+1\right)  }e^{\phi}$. By
replacing in the field equations we end with the dynamical system
\cite{ch1,ch2}%
\begin{equation}
3H^{2}-\frac{\lambda}{2N^{2}}\dot{\phi}^{2}-V\left(  \phi\right)  -\tilde
{\rho}_{m0}a^{-3\left(  w_{m}+1\right)  }e^{\frac{\phi}{2}}=0, \label{c.20}%
\end{equation}%
\begin{equation}
\dot{H}+H^{2}+\frac{1}{6}\tilde{\rho}_{m0}a^{-3\left(  w_{m}+1\right)
}e^{\frac{\phi}{2}}\left(  1+w_{m}\right)  +\frac{1}{3}\left(  \lambda
\dot{\phi}^{2}-V\left(  \phi\right)  \right)  =0, \label{c.21}%
\end{equation}%
\begin{equation}
\ddot{\phi}+3H\dot{\phi}+V\left(  \phi\right)  +\frac{1}{2\lambda}\tilde{\rho
}_{m0}a^{-3\left(  w_{m}+1\right)  }e^{\frac{\phi}{2}}=0, \label{c.22}%
\end{equation}

\bigskip The latter field equations follow from the variation of the
point-like Lagrangian function%
\begin{equation}
\mathcal{L}\left(  a,\dot{a},\phi,\dot{\phi}\right)  =\left(  -3a\dot{a}%
^{2}+\frac{\lambda}{2}a^{3}\dot{\phi}^{2}\right)  -\left(  a^{3}V\left(
\phi\right)  +\tilde{\rho}_{m0}e^{\frac{\phi}{2}}a^{w_{m}}\right)  .
\label{ee.06}%
\end{equation}
Specifically, equations (\ref{c.21}) and (\ref{c.22}) follow from the
variation of Lagrangian (\ref{ee.06}) with respect to the dynamical variables
$a$ and $\phi$, while equation (\ref{c.20}) can be seen as the energy
constraint for the two-dimensional conserved system.

\section{Analytic Solution}

\label{sec3}

Dark matter is described by a pressureless fluid source, that is, $w_{m} = 0$.
Furthermore, for the scalar field potential, we consider the exponential
function $V\left(  \phi\right)  = V_{0} \exp\left(  \phi\right)  $. For this
cosmological model, the field equations (\ref{c.20}), (\ref{c.21}), and
(\ref{c.22}) admit the following conservation law \cite{weylnoether}:
\begin{equation}
I_{0}=-4a^{2}\dot{a}-4\lambda a^{3}\dot{\phi}. \label{ee.07}%
\end{equation}
This conservation law follows from the existence of a nontrivial variational
symmetry for the Lagrangian (\ref{ee.06}).

Under the change of variables $\psi=\phi+6\ln a$, the conservation law
(\ref{ee.07}) reads $I_{0}=\frac{2}{3}ap_{a},$ while the point-like Lagrangian
(\ref{ee.06}) is given by the following expression%
\begin{equation}
\mathcal{L}\left(  a,\dot{a},\psi,\dot{\psi}\right)  =3\left(  1-6\lambda
\right)  a\dot{a}^{2}+6\lambda a^{2}\dot{a}\dot{\psi}-\frac{1}{2}\lambda
a^{3}\dot{\psi}^{2}+a^{-3}\left(  \tilde{\rho}_{m0}e^{\frac{\psi}{2}}%
+V_{0}e^{\psi}\right)  .
\end{equation}

The solution of the field equations can be easily recovered using Hamilton's
Jacobi theory. We introduce the momenta $p_{a}=\frac{\partial\mathcal{L}%
}{\partial\dot{a}}$, and $p_{\psi}=\frac{\partial\mathcal{L}}{\partial
\dot{\psi}}$. Thus, in the Hamiltonian formalism, the cosmological model is
described by the function
\begin{equation}
\mathcal{H}\left(  a,\psi,p_{a},p_{\psi}\right)  \equiv\frac{1}{12a}p_{a}%
^{2}+\frac{1}{a^{2}}p_{a}p_{\psi}+\frac{\left(  6\lambda-1\right)  }{2\lambda
a^{3}}p_{\psi}^{2}-a^{-3}\left(  \tilde{\rho}_{m0}e^{\frac{\psi}{2}}%
+V_{0}e^{\psi}\right)  =0.
\end{equation}
Thus, from the Hamilton-Jacobi equation we determine the Action%
\begin{equation}
S\left(  a,\psi\right)  =\frac{3}{2}I_{0}\ln a+\int\frac{\pm\sqrt
{\kappa\left(  \psi\right)  }-6I_{0}\lambda}{4e\left(  6\lambda-1\right)
}d\psi
\end{equation}
with%
\begin{equation}
\kappa\left(  \psi\right)  =6I_{0}^{2}\lambda+32\lambda\left(  6\lambda
-1\right)  \left(  V_{0}e^{\psi}+\tilde{\rho}_{m0}e^{\frac{\psi}{2}}\right)  .
\end{equation}

Therefore, the cosmological solution is expressed by the reduced system%
\begin{equation}
H\left(  a\right)  =\frac{\pm\sqrt{\kappa\left(  \psi\left(  a\right)
\right)  }-I_{0}}{4a^{3}\left(  6\lambda-1\right)  },
\end{equation}
where
\begin{equation}
\frac{d\psi}{da}=\frac{\left(  6\lambda-1\right)  \sqrt{\kappa\left(
\psi\left(  a\right)  \right)  }}{a\lambda\left(  \sqrt{\kappa\left(
\psi\left(  a\right)  \right)  }\mp I_{0}\right)  }. \label{ee.09}%
\end{equation}

We observe that, in order to avoid singularities, the parameter $I_{0}$ is
constrained as $\mp I_{0} > 0$. Moreover, for $\lambda> \frac{1}{6}$, it
follows from (\ref{ee.09}) that $\frac{d\psi}{da} > 0$. Therefore,
$\kappa\left(  \psi\right)  \simeq6 I_{0}^{2} \lambda+ 32 \lambda\left(
6\lambda- 1 \right)  \left(  V_{0} e^{\psi} + \tilde{\rho}_{m0} e^{\frac{\psi
}{2}} \right)  $, and the evolution of the scalar field $\psi$ is very quickly
asymptotically described by the differential equation $\frac{d\psi}{da}
\simeq\frac{\left(  6\lambda- 1 \right)  }{a\lambda}$. Therefore, the analytic
expression for the Hubble function takes the following form:
\begin{equation}
H_{WIS}\left(  a\right)  =H_{0}\sqrt{\left(  1-\tilde{\Omega}_{m0}\right)
a^{-w_{0}}+\tilde{\Omega}_{m0}a^{-3-\frac{1}{2}w_{0}}},~w_{0}=\lambda^{-1}
\label{ee.10}%
\end{equation}
where ,~$H_{0}^{2}\tilde{\Omega}_{m0}=\frac{2\lambda}{6\lambda-1}\tilde{\rho
}_{m0}$ and $H_{0}^{2}\left(  1-\tilde{\Omega}_{m0}\right)  =\frac{2\lambda
}{6\lambda-1}V_{0}$. In the limit $w_{0}\rightarrow0$, i.e. $\lambda
\rightarrow\infty$, the $\Lambda$CDM is recovered~and $\tilde{\Omega}%
_{m0}\rightarrow\Omega_{m0}$, that is,
\begin{equation}
H_{\Lambda}\left(  a\right)  =H_{0}\sqrt{\left(  1-\Omega_{m0}\right)
+\Omega_{m0}a^{-3}}. \label{ee.11}%
\end{equation}
We remark that the existence of the coupling parameter $\lambda$ leads to a
deviation between the two components of the cosmological fluid due to the
nonzero interaction term.

For $\lambda> \frac{1}{6}$, the parameter $I_{0}$ is not an essential constant
for the late-time solution. On the other hand, for $\lambda< \frac{1}{6}$, it
is possible for $\kappa\left(  \psi\right)  < 0$, which may lead to singular
behavior except in the case where $V_{0} < 0$ and $\tilde{\rho}{m0} < 0$.
However, in this latter scenario, the $\Lambda$CDM limit is not recovered.
Therefore, to overcome this limitation, we consider $I{0} = 0$.

\section{Observational Constraints}

\label{sec4}

In this section, we use late-time cosmological observations to constrain the
viability of the cosmological model (\ref{ee.10}) and to probe its deviation
from $\Lambda$CDM due to the presence of the interaction term.

\subsection{Cosmological data}

In order to constrain our model, we make use of the following cosmological observations.

\begin{itemize}
\item Supernova (SNIa): We consider the Pantheon+ catalogue, includes 1701
light curves of 1550 spectroscopically confirmed supernova events within the
range $10^{-3}<z<2.27~$\cite{pan}. We consider the Pantheon+ data without the
Cepheid calibration. The luminosity distance~$D_{L}=c\left(  1+z\right)
\int\frac{dz}{H\left(  z\right)  }$, is used to define the theoretical
distance modulus$~\mu^{th}=5\log D_{L}+25$, which is used to constraint with
the distance modulus $\mu^{obs}~$at~observed redshifts~$z$.

\item Observational Hubble Data (OHD): These data include direct measurements
of the Hubble parameter, without any cosmological assumptions, obtained from
the differential age evolution of cosmic chronometers. Cosmic chronometers are
old, passively evolving galaxies with synchronous stellar populations and
similar cosmic evolution \cite{co01}. This data set is model independent. We
make use of the 31 direct measurements of the Hubble parameter for redshifts
in the range $0.09\leq z\leq1.965$~\cite{cc1}.

\item Baryonic acoustic oscillations (BAO): These data are data from the DESI
DR2 release \cite{des4,des5,des6}. The dataset includes observation values of
the transverse comoving angular distance ratio $\frac{D_{M}}{r_{d}}%
=\frac{\left(  1+z\right)  ^{-1}D_{L}}{r_{d}},$ the volume averaged distance
ratio$~\frac{D_{V}}{r_{d}}=\frac{\left(  cD_{L}\frac{z}{H\left(  z\right)
}\right)  ^{\frac{1}{3}}}{r_{d}}$~and the Hubble distance ratio$~\frac{D_{H}%
}{r_{d}}=\frac{c}{r_{d}H}$ where $r_{d}$ is the sound horizon at the drag epoch.

\item Gamma-ray bursts (GRB): \ We consider the 193 events where they have
with the use of the Amati correlation \cite{amm} in the redshift range
$0.0335<z<8.1$. The distance modulus $\mu^{obs}$ for each event at the
observed redshift are summarized in \cite{amti}. The Amati correlation is a
model-independent approach which allows to use GRBs as distance indicators.
\end{itemize}

We use different combinations of the above data. Specifically we consider the
datasets~$\mathbf{D}_{1}$,~$\mathbf{D}_{2}$, $\mathbf{D}_{3}$ and
$\mathbf{D}_{4}$. The datasets are defined as $\mathbf{D}_{1}:~$SNIa \& BAO;
$\mathbf{D}_{2}:~$SNIa~\& OHD~\& BAO; $\mathbf{D}_{3}:$SNIa \& BAO \& GRBs and
$\mathbf{D}_{4}:$ SNIa \& OHD \& BAO \& GRBs. These datasets are described in
Table \ref{table0a}.%

\begin{table}[tbp] \centering
\caption{Datasets Employed for Observational Constraints.}%
\begin{tabular}
[c]{ccccc}\hline\hline
\textbf{Dataset} & \textbf{SNIa} & \textbf{OHD} & \textbf{BAO} &
\textbf{GRBs}\\\hline
$\mathbf{D}_{1}$ & $\checkmark$ & $\times$ & $\checkmark$ & $\times$\\
$\mathbf{D}_{2}$ & $\checkmark$ & $\checkmark$ & $\checkmark$ & $\times$\\
$\mathbf{D}_{3}$ & $\checkmark$ & $\times$ & $\checkmark$ & $\checkmark$\\
$\mathbf{D}_{4}$ & $\checkmark$ & $\checkmark$ & $\checkmark$ & $\checkmark
$\\\hline\hline
\end{tabular}
\label{table0a}%
\end{table}%

\subsection{Methodology}

To perform our statistical analysis by using the above datasets we use the
Bayesian inference COBAYA\footnote{https://cobaya.readthedocs.io/}
\cite{cob1,cob2} with the PolyChord nested sampler \cite{poly1,poly2}, which
provides the Bayesian evidence, an important information for the statistical
comparison between models with different degrees of freedom.

We constrain the Hubble function provided by WIS within the four-dimensional
parameter space $\left\{  H_{0}, \hat{\Omega}_{m0}, w_{0}, r_{d} \right\}  $,
and the Hubble function of the $\Lambda$CDM model within the three-dimensional
parameter space $\left\{  H_{0}, \Omega_{m0}, r_{d} \right\}  $. The parameter
$\hat{\Omega}_{m0}$ describes the effective energy density of dark matter and
baryons. For the datasets considered, the effects of radiation can be neglected.

Since WIS and $\Lambda$CDM have different dimensions in parameter space, we
compare them using the Akaike Information Criterion (AIC) \cite{AIC} and
Jeffrey's scale \cite{AIC2}.

From the minimum $\chi^{2}=-2\ln\mathcal{L}_{\max}$ value of each model we
define the $AIC$ parameter using the expression \cite{AIC}
\begin{equation}
AIC\simeq-2\ln\mathcal{L}_{\max}+2\kappa.
\end{equation}
where $\kappa$ is the dimension of the parametric space. For the WIS
$\kappa=4$ while for the $\Lambda$CDM $\kappa=3$. The difference of the $AIC$
parameters reads%
\[
\Delta AIC=AIC_{WIC}-AIC_{\Lambda}=-2\ln\frac{\mathcal{L}_{\max}^{WIC}%
}{\mathcal{L}_{\max}^{\Lambda}}+2.
\]
Thus, according to the Akaike's scale for $\left\vert \Delta AIC\right\vert
<2$, the two models are statistical equivalent, $\left\vert \Delta
AIC\right\vert <6$ there is a weak evidence in favor of the model with smaller
$AIC$,~if $\left\vert \Delta AIC\right\vert >6$ there is a strong evidence,
while for $\left\vert \Delta AIC\right\vert >10$ there exist a clear evidence
for the preference of the model with lover $AIC$ value.

Jeffrey's scale \cite{AIC2} is defined by the difference of the Bayesian
evidences $\ln Z_{WIC}$ and $\ln Z_{\Lambda}$.

Let
\[
\Delta\left(  \ln Z\right)  =\ln\frac{Z_{WIC}}{Z_{\Lambda}}%
\]
be the difference between the Bayesian evidence values for the two models.
Then, if $\left\vert \Delta\left(  \ln Z\right)  \right\vert \ll1$, the two
models are comparable, and there is no evidence favoring either model. For
$\left\vert \Delta\left(  \ln Z\right)  \right\vert <1$, there is weak
evidence in favor of the model with the higher Bayesian evidence value. When
$\left\vert \Delta\left(  \ln Z\right)  \right\vert >1$, the evidence is
moderate; when $\left\vert \Delta\left(  \ln Z\right)  \right\vert >2.5$, the
evidence is strong; and when $\left\vert \Delta\left(  \ln Z\right)
\right\vert >5$, there is very strong evidence that the model with the higher
value, $\ln Z_{2}$, is statistically preferred.

The above criteria for the statistical comparison of the models are summarized
in\ Table \ref{table0}.%

\begin{table}[tbp] \centering
\caption{Akaike's and Jeffrey's scales.}%
\begin{tabular}
[c]{cccc}\hline\hline
$\mathbf{\Delta}\left(  AIC\right)  $ & \textbf{Akaike's scale} &
$\Delta\left(  \ln Z\right)  $ & \textbf{Jeffrey's scale}\\\hline
$<2$ & Inconclusive & $<<1$ & Inconclusive\\
$2-6$ & Weak evidence & $<1$ & Weak evidence\\
$6-10$ & Strong evidence & $1-2.5$ & Moderate evidence\\
$>10$ & Clear evidence & $>5$ & Clear evidence\\\hline\hline
\end{tabular}
\label{table0}%
\end{table}%

\subsection{Results}

We constrain the Hubble function of the WIS model (\ref{ee.10}) and that of
the $\Lambda$CDM model (\ref{ee.11}) using the four different datasets defined
in Table~\ref{table0a}.

For the dataset $\mathbf{D}{1}$, the free parameters of the WIS model are
constrained as $H{0} = 68.3_{-4.3}^{+3.3}$, $\tilde{\Omega}{m0} =
0.224{-0.064}^{+0.027}$, and $w_{0} = 0.46_{-0.17}^{+0.36}$. By comparing the
statistical indicators with those of the $\Lambda$CDM model, we find
$\mathbf{\chi}{\min,\text{WIS}}^{2} - \mathbf{\chi}{\min,\Lambda}^{2} = -2.4$,
which indicates that WIS provides a better fit to the data. However, the AIC
yields $\Delta\left(  \text{AIC} \right)  = -0.4$, and the Bayesian evidence
gives $\Delta\left(  \ln Z \right)  = +0.67$. According to Table~\ref{table0},
Akaike's scale indicates that the two models fit the data equally well, while
Jeffrey's scale suggests weak evidence in favor of WIS.

The introduction of the OHD data in $\mathbf{D}{2}$ modifies the free
parameters to $H{0} = 68.1_{-1.7}^{+1.7}$, $\tilde{\Omega}{m0} =
0.229{-0.067}^{+0.29}$, and $w_{0} = 0.44_{-0.19}^{+0.36}$. The statistical
indicators are $\mathbf{\chi}{\min,\text{WIS}}^{2} - \mathbf{\chi}%
{\min,\Lambda}^{2} = -2.1$, $\Delta\left(  \text{AIC} \right)  = -0.1$, and
$\Delta\left(  \ln Z \right)  = +0.52$. Therefore, the conclusions remain the
same as in the previous case.

For the dataset $\mathbf{D}_{3}$ we calculate $H_{0}=68.8_{-3.4}^{+3.4}%
$,~$\tilde{\Omega}_{m0}=0.199_{-0.043}^{+0.021}$ and $w_{0}=0.55_{-0.15}%
^{+0.25}$ and the comparison with the $\Lambda$CDM gives $\mathbf{\chi}_{\min
WIS}^{2}-\mathbf{\chi}_{\min\Lambda}^{2}=-4.8,$ $\Delta\left(  AIC\right)
=-2.8$ and $\Delta\left(  \ln Z\right)  =+1.48$. From the AIC we conclude that
there exist a weak evidence in favor of WIS, while Jeffrey's scale states that
there exists a moderate evidence in favor of WIS.

Finally, the combination of all the data in dataset $\mathbf{D}_{4}$ leads to
the cosmological parameters~$H_{0}=68.6_{-1.8}^{+1.6}$,~$\tilde{\Omega}%
_{m0}=0.206_{-0.048}^{+0.023}$ and $w_{0}=0.51_{-0.15}^{+0.27}$,
$\mathbf{\chi}_{\min WIS}^{2}-\mathbf{\chi}_{\min\Lambda}^{2}=-4.0,$
$\Delta\left(  AIC\right)  =-2.0$ and $\Delta\left(  \ln Z\right)  =+1.28$
where we make the same conclusions with that of dataset $\mathbf{D}_{3}$.

\begin{figure}[ptbh]
\centering\includegraphics[width=0.7\textwidth]{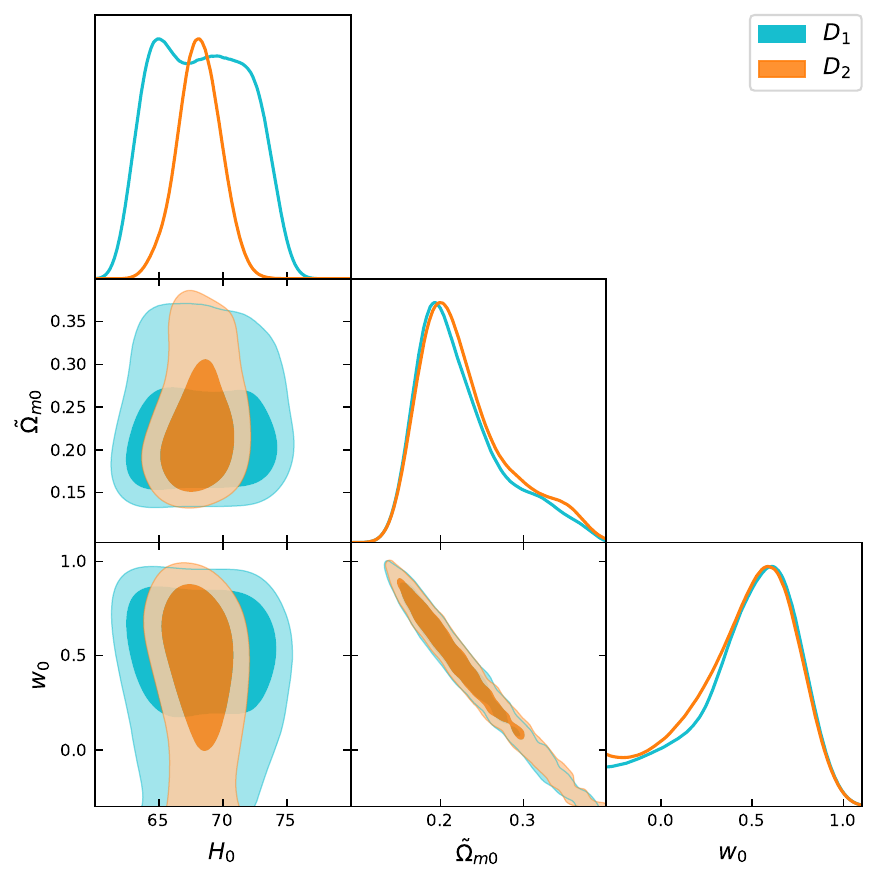}\caption{Confidence
space for the best-fit parameters for WIS Hubble function (\ref{ee.10}) for
the datasets $D_{1}$ and $D_{2}$.}%
\label{fig1}%
\end{figure}

\begin{figure}[ptbh]
\centering\includegraphics[width=0.7\textwidth]{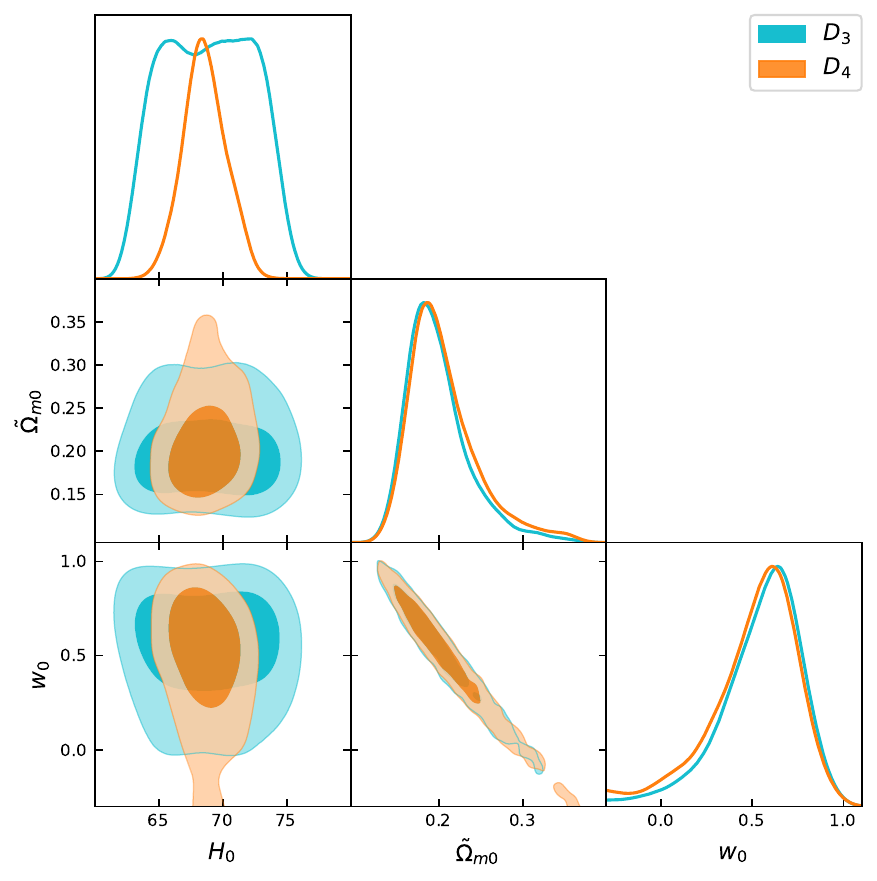}\caption{Confidence
space for the best-fit parameters for WIS Hubble function (\ref{ee.10}) for
the datasets $D_{3}$ and $D_{4}$.}%
\label{fig2}%
\end{figure}

The best values for the free parameters and the statistical comparison with
$\Lambda$CDM are presented in Tables \ref{table1} and \ref{table2}. In Figs.
\ref{fig1} and \ref{fig2} we present the confidence space for the best-fit
parameters of the WIS model.

From this analysis, we observe that the inclusion of GRBs in the dataset leads
to a preference for the WIS model over $\Lambda$CDM. However, this preference
becomes weaker when the OHD data are also included. It is important to note
that, for all datasets, we find that the value $w_{0} = 0$ 'which corresponds
to the $\Lambda$CDM limit is included within the $2\sigma$ confidence level.

Last but not least, a positive value of $w_{0}$ is supported by the data. That
indicates that $\lambda> 0$, from which we conclude that the scalar field is
of the quintessence type.

\begin{table}[tbp] \centering
\caption{Obeservation Constraints and Best Fit Parameters for WIS.}%
\begin{tabular}
[c]{ccccc}\hline\hline
\textbf{Dataset} & $\mathbf{D}_{1}$ & $\mathbf{D}_{2}$ & $\mathbf{D}_{3}$ &
$\mathbf{D}_{4}$\\
$\mathbf{H}_{0}$ & $68.3_{-4.3}^{+3.3}$ & $68.1_{-1.7}^{+1.7}$ &
$68.8_{-3.4}^{+3.4}$ & $68.6_{-1.8}^{+1.6}$\\
$\mathbf{\tilde{\Omega}}_{m0}$ & $0.224_{-0.064}^{+0.027}$ & $0.229_{-0.067}%
^{+0.29}$ & $0.199_{-0.043}^{+0.021}$ & $0.206_{-0.048}^{+0.023}$\\
$w_{0}$ & $0.46_{-0.17}^{+0.36}$ & $0.44_{-0.19}^{+0.36}$ & $0.55_{-0.15}%
^{+0.25}$ & $0.51_{-0.15}^{+0.27}$\\\hline\hline
\end{tabular}
\label{table1}%
\end{table}%
%

\begin{table}[tbp] \centering
\caption{Statistical comparison of WIS with the $\Lambda$CDM}%
\begin{tabular}
[c]{cccccc}\hline\hline
\textbf{Dataset} & $\Delta\left(  \mathbf{\chi}_{\min WIS-\Lambda}^{2}\right)
$ & \textbf{AIC}$_{WIS}-$\textbf{AIC}$_{\Lambda}$ & \textbf{Akaike's scale} &
$\log\frac{\mathbf{Z}_{WIS}}{\boldsymbol{Z}_{\Lambda}}$ & \textbf{Jeffrey's
scale}\\
$\mathbf{D}_{1}$ & $-2.4$ & $-0.4$ & Incoclusive & $+0.67$ & Weak evidence for
WIS\\
$\mathbf{D}_{2}$ & $-2.1$ & $-0.1$ & Incoclusive & $+0.52$ & Weak evidence for
WIS\\
$\mathbf{D}_{3}$ & $-4.8$ & $-2.8$ & Weak evidence for WIS & $+1.48$ &
Moderate evidence for WIS\\
$\mathbf{D}_{4}$ & $-4.0$ & \thinspace\thinspace$-2.0$ & Weak evidence for
WIS & $+1.28$ & Moderate evidence for WIS\\\hline\hline
\end{tabular}
\label{table2}%
\end{table}%

\section{Conclusions}

\label{sec5}

We examined the cosmological viability of the WIS in a spatially flat universe
by constraining an analytic model with observational data. Within the WIS, the
scalar field arises from the existence of the nonmetricity component of the
connection. Furthermore, when matter is introduced, there exists a coupling
function that introduces energy transfer between the scalar field and the dark
matter. This interaction term indicates the existence of a Chameleon mechanism
for the scalar field.

The field equations for this gravitational model form a constrained
Hamiltonian system of two second-order ordinary differential equations. For
the exponential scalar field potential, the field equations are integrable
according to Liouville, because they admit a nontrivial Noetherian
conservation law. This allows us to reduce the field equations into a system
of two first-order differential equations. We investigate the nature of all
the constants in the solution and find that, for a late-time cosmologically
viable model, one of the parameters can be omitted. Thus, we end up with an
analytic expression for the Hubble function which depends on one additional
free parameter compared to the $\Lambda$CDM. When this additional parameter
vanishes, $\Lambda$CDM is recovered.

Using the Pantheon+ Supernova data, the Cosmic Chronometers, the DESI DR2 BAO
data, and the GRBs, we constrain the free parameters of the WIS and test it
against the $\Lambda$CDM. We find that, from these data, there is a preference
for a quintessence-like scalar field $\left(  \lambda>0\right)  $, while the
two models differ at the $2\sigma$ level. Last but not least, $\Lambda$CDM is
challenged by the WIS model, because not only does it constrain the data in a
better way consistently providing a larger likelihood value, but also the
statistical criterion of Jeffrey's scale based on the Bayesian evidence
indicates that there is weak evidence in favor of WIS, which becomes moderate
when the GRBs dataset is included in the analysis.

Finally, we calculated smaller values for the Hubble constant in comparison
with the $\Lambda$CDM. This statement remains true even when we introduce the
SH0ES calibration of the Supernova data, which means that WIS may be able to
provide a solution to the $H_{0}$ tension. However, such analysis will be
published elsewhere.


\begin{acknowledgments}
The author thanks the support of Vicerrector\'{\i}a de Investigaci\'{o}n y
Desarrollo Tecnol\'{o}gico (Vridt) at Universidad Cat\'{o}lica del Norte
through N\'{u}cleo de Investigaci\'{o}n Geometr\'{\i}a Diferencial y
Aplicaciones, Resoluci\'{o}n Vridt No - 096/2022. This work was partially
supported by Proyecto Fondecyt Regular Folio 1240514, Etapa 2025.
\end{acknowledgments}

\end{document}